\documentclass[aps,prl,twocolumn,superscriptaddress,showpacs,preprintnumbers,amsmath,amssymb]{revtex4}

\usepackage{graphicx} 
\usepackage{dcolumn}  

\newcommand{\MeV}{\mathrm{Me\kern -0.1em V}}
\newcommand{\GeV}{\mathrm{Ge\kern -0.1em V}}
\newcommand{\TeV}{\mathrm{Te\kern -0.1em V}}

\newcommand{\EE}{\ensuremath{{e}^+{e}^-}}

\newcommand{\RA}{\ensuremath{\rightarrow}}

\newcommand{\eegg}{\EE\ensuremath{\rightarrow \gamma \gamma}}
\newcommand{\bgg}{\ensuremath{B^0 \RA \gamma \gamma}}
\newcommand{\bpipi}{\ensuremath{B^0 \RA \pi^0 \pi^0}}

\newcommand{\betapi}{\ensuremath{B^0 \RA \eta \pi^0}}
\newcommand{\bbar}{\ensuremath{B\overline{B}}}
\newcommand{\invfb}{\ensuremath{\mathrm{fb^{-1}}}}
\newcommand{\mbc}{\ensuremath{M_{\mathrm{bc}}}}
\newcommand{\deltae}{\ensuremath{\Delta E}}
\newcommand{\costb}{\ensuremath{\cos \theta^*_B}}
\newcommand{\etal}{{\it et al.}}

\newcommand{\PL}     {Phys.\ Lett.\ }
\newcommand{\PR}     {Phys.\ Rev.\ }
\newcommand{\PRL}    {Phys.\ Rev.\ Lett.\ }

\begin{document}



\vspace*{2\baselineskip}

\title{  \quad\\[0.5cm]\boldmath Search for the decay \bgg}

\affiliation{Budker Institute of Nuclear Physics, Novosibirsk}
\affiliation{Chonnam National University, Kwangju}
\affiliation{University of Cincinnati, Cincinnati, Ohio 45221}
\affiliation{University of Hawaii, Honolulu, Hawaii 96822}
\affiliation{High Energy Accelerator Research Organization (KEK), Tsukuba}
\affiliation{Institute of High Energy Physics, Chinese Academy of Sciences, Beijing}
\affiliation{Institute of High Energy Physics, Vienna}
\affiliation{Institute of High Energy Physics, Protvino}
\affiliation{Institute for Theoretical and Experimental Physics, Moscow}
\affiliation{J. Stefan Institute, Ljubljana}
\affiliation{Kanagawa University, Yokohama}
\affiliation{Korea University, Seoul}
\affiliation{Swiss Federal Institute of Technology of Lausanne, EPFL, Lausanne}
\affiliation{University of Ljubljana, Ljubljana}
\affiliation{University of Maribor, Maribor}
\affiliation{University of Melbourne, Victoria}
\affiliation{Nagoya University, Nagoya}
\affiliation{Nara Women's University, Nara}
\affiliation{National Central University, Chung-li}
\affiliation{National United University, Miao Li}
\affiliation{Department of Physics, National Taiwan University, Taipei}
\affiliation{H. Niewodniczanski Institute of Nuclear Physics, Krakow}
\affiliation{Niigata University, Niigata}
\affiliation{Nova Gorica Polytechnic, Nova Gorica}
\affiliation{Osaka City University, Osaka}
\affiliation{Osaka University, Osaka}
\affiliation{Panjab University, Chandigarh}
\affiliation{Peking University, Beijing}
\affiliation{Princeton University, Princeton, New Jersey 08544}
\affiliation{RIKEN BNL Research Center, Upton, New York 11973}
\affiliation{University of Science and Technology of China, Hefei}
\affiliation{Seoul National University, Seoul}
\affiliation{Shinshu University, Nagano}
\affiliation{Sungkyunkwan University, Suwon}
\affiliation{University of Sydney, Sydney NSW}
\affiliation{Tata Institute of Fundamental Research, Bombay}
\affiliation{Toho University, Funabashi}
\affiliation{Tohoku Gakuin University, Tagajo}
\affiliation{Tohoku University, Sendai}
\affiliation{Department of Physics, University of Tokyo, Tokyo}
\affiliation{Tokyo Institute of Technology, Tokyo}
\affiliation{Tokyo Metropolitan University, Tokyo}
\affiliation{University of Tsukuba, Tsukuba}
\affiliation{Virginia Polytechnic Institute and State University, Blacksburg, Virginia 24061}
\affiliation{Yonsei University, Seoul}
   \author{S.~Villa}\affiliation{Swiss Federal Institute of Technology of Lausanne, EPFL, Lausanne} 
   \author{K.~Abe}\affiliation{High Energy Accelerator Research Organization (KEK), Tsukuba} 
   \author{K.~Abe}\affiliation{Tohoku Gakuin University, Tagajo} 
   \author{I.~Adachi}\affiliation{High Energy Accelerator Research Organization (KEK), Tsukuba} 
   \author{H.~Aihara}\affiliation{Department of Physics, University of Tokyo, Tokyo} 
   \author{Y.~Asano}\affiliation{University of Tsukuba, Tsukuba} 
   \author{T.~Aushev}\affiliation{Institute for Theoretical and Experimental Physics, Moscow} 
   \author{S.~Bahinipati}\affiliation{University of Cincinnati, Cincinnati, Ohio 45221} 
   \author{A.~M.~Bakich}\affiliation{University of Sydney, Sydney NSW} 
   \author{V.~Balagura}\affiliation{Institute for Theoretical and Experimental Physics, Moscow} 
% ...
\collaboration{The Belle Collaboration}
\noaffiliation

\begin{abstract}
The rare decay  \bgg \ is searched for in 104 fb$^{-1}$ of data, 
corresponding to $111 \times 10^6$ $\bbar$ pairs, collected 
with the Belle detector at the KEKB asymmetric-energy $e^+ e^-$
collider.
No evidence for the signal is found, and an upper
limit of $6.2 \times 10^{-7}$ at 90\% confidence level is set for the corresponding branching fraction.
\end{abstract}


\maketitle

\begin{figure}[htb]
\includegraphics[width=0.4\textwidth]{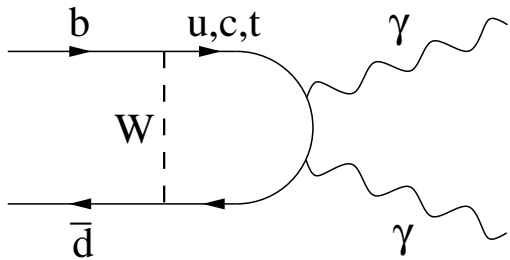}
\caption{A possible diagram contributing to \bgg \ at the 
lowest order in the SM. The exchange of a charged Higgs boson instead
of the $W$ boson could contribute to this process in some extensions
of the SM.}
\label{fig1}
\end{figure}

{\renewcommand{\thefootnote}{\fnsymbol{footnote}}}
\setcounter{footnote}{0}
%
The channel \bgg \ is a rare decay of the $B^0$ meson that is interesting both
experimentally, for its remarkably clean signature, and 
theoretically, as a tool for constraining physics beyond
the Standard Model (SM).
The SM prediction for the \bgg \ branching fraction (BF) is around
$3 \times 10^{-8}$~\cite{SMpredictions}, and
a possible Feynman diagram contributing to this channel 
is shown in Fig.~\ref{fig1}.
Sizable enhancements of the BF are predicted in many new physics 
models~\cite{newphysics}; a typical contribution arising from non-SM
effects would follow from the replacement of the $W$ boson in Fig.~\ref{fig1} 
with another charged particle such as a charged Higgs boson.
The \bgg \ channel is also interesting because
it allows the study of non-trivial QCD dynamics in $B$ decay, 
via a pure non-hadronic final state.

Experimental limits on the BF have been
set by L3~\cite{l3bgg} and BaBar~\cite{babarbgg}.
The BaBar upper limit of
$1.7 \times 10^{-6}$ at 90\% confidence level
(CL), obtained with 19.4 \invfb \ of data, is 
the most restrictive existing experimental constraint on this channel.

The present search for the  \bgg \ decay is based on a data sample 
of 104 \invfb, which
contains $111 \times 10^6$  \bbar \ pairs, 
collected  with the Belle detector at the KEKB asymmetric-energy
$e^+e^-$ (3.5 on 8~GeV) collider~\cite{KEKB}
operating at the $\Upsilon(4S)$ resonance.

The Belle detector is a large-solid-angle magnetic
spectrometer that
consists of a 4-layer silicon vertex detector,
a small-cell inner drift chamber~\cite{Ushiroda},
a 47-layer central drift chamber, an array of
aerogel threshold \v{C}erenkov counters, 
a barrel-like arrangement of time-of-flight
scintillation counters, and an electromagnetic calorimeter
comprised of CsI(Tl) crystals located inside 
a super-conducting solenoid coil that provides a 1.5~T
magnetic field.  An iron flux-return located outside of
the coil is instrumented to detect $K_L^0$ mesons and to identify
muons.  The detector
is described in detail elsewhere~\cite{Belle}.

%
%

The \bgg \ events are characterized in the center-of-mass (CM) frame
by two back-to-back highly energetic photons. 
Photons are selected from isolated clusters in the calorimeter
that are not matched to charged tracks. 
We require a shower shape consistent with that of a photon: 
for each cluster,  
the ratio of the energy deposited in the array of the central 
$3\times 3$ calorimeter cells to that of $5\times 5$ cells is
computed, and clusters with a ratio smaller than 0.95 are rejected. 

In the Belle detector, a large background for this channel 
is due to the overlap of a hadronic event
with energy deposits left in the calorimeter by previous 
QED interactions (mainly Bhabha scattering). 
Such composite events are completely removed
using timing information for calorimeter clusters 
associated with the candidate photons. 
Only photons that are in time with the rest of the event are 
retained. The efficiency of this selection on signal events
is larger than 99.5\%.
The cluster timing information is stored in the raw data,
and is available in the reduced format used for
analyses only for data processed after the summer of 2004, 
thus limiting the dataset 
available for this analysis to 104 \invfb.  

Rejection of $\pi^0$ and $\eta$ mesons is of primary importance in
a search for a purely radiative rare decay of the $B^0$ meson.
All pairs of photons with 
energy larger than 50 $\MeV$  and a maximum absolute value of the 
difference between their invariant 
mass and the $\pi^0$ mass~\cite{PDG} of 15 $\MeV/c^2$ are
identified as neutral pions.
For $\eta$ mesons, the minimum energy requirement is 100 $\MeV$, and the invariant mass
of the two photons is required to be within 60 $\MeV/c^2$ of the $\eta$
mass~\cite{PDG}.  All pairs of photons passing either the $\pi^0$ or $\eta$ selection are
removed from subsequent analysis.

The two highest-energy photons are selected and
their momenta are added to reconstruct the $B^0$ momentum. 
Two variables are used to separate signal events from background:
$\mbc =\sqrt{E^{*2}_{\mathrm{beam}}/c^4 - p^{*2}_{B}/c^2}$ and
$\deltae = E^*_B - E^*_{\mathrm{beam}}$, where
$E^{*}_{\mathrm{beam}}$ is the beam energy and $E^*_B$ and $p^{*}_{B}$ are the reconstructed
$B^0$ energy and momentum, all variables being evaluated in the CM frame.
The signal is concentrated near $\deltae = 0$ and \mbc \ equal to the
$B^0$ mass. The signal window is therefore defined as
\begin{eqnarray}\nonumber 
-0.25\; \GeV & < & \deltae < 0.15\; \GeV  \\ \nonumber
5.272\; \GeV/c^2 & < & \mbc < 5.288\; \GeV/c^2\; 
\end{eqnarray}
corresponding to about two standard deviation intervals above and
below the central values just mentioned.

The main background for the \bgg \ channel is due to continuum events, mostly 
coming from light quark pair
production and fragmentation ($u\bar{u}$, $d\bar{d}$, and $s\bar{s}$, uds for short).
Two variables that display quite powerful separation between signal
and continuum background are
a Fisher discriminant based on modified Fox-Wolfram moments~\cite{SFW} 
and the $B^0$ production
angle with respect to the beam in the CM frame, \costb.
These variables are combined in a likelihood ratio, LR;
signal and background distributions used to construct the
LR are extracted from Monte Carlo (MC) samples.
In the continuum background, the two particles 
that are reconstructed as
photons are more abundantly produced at low polar angle ($\theta^*$, measured
in the CM frame), while the signal
photons have a flat distribution in $\cos \theta^*$.
Selection requirements on LR  (LR $> 0.92$)
and on the cosine of the polar angle of the most energetic
photon  in the event ($| \cos\theta^* |< 0.65$) are optimized 
by maximizing  $N_{\mathrm{sig}} / \sqrt{N_{\mathrm{sig}}+ N_{\mathrm{bck}}} $, 
where $N_{\mathrm{sig}}$ ($N_{\mathrm{bck}}$) is the expected number
of signal (background) events in the signal window. 
The expected numbers of events are computed for an integrated 
luminosity of 104 \invfb \ and assuming for the signal the 
BF predicted by the SM and for the background the prediction
of the continuum MC.
The above requirements reduce the continuum background in the signal
window by a factor of 55, while retaining 31\% of signal events.

\begin{figure}[htb]
\includegraphics[width=0.4\textwidth]{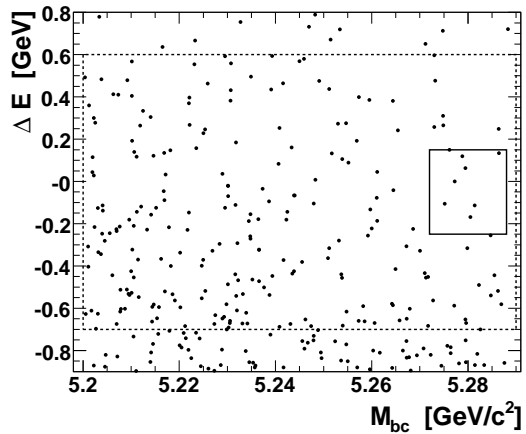}
\caption{\deltae \ versus \mbc \ for data events, selected as described in
the text. Seven events are in the signal window (solid rectangle). The
fit window is shown as a dashed rectangle.}
\label{fig2}
\end{figure}

The total selection efficiency for signal, 
evaluated using MC events, is 11.7\%.
In data, seven events lie in the signal window.
They are shown in the \deltae--\mbc \ 
plane in Fig.~\ref{fig2}, where the
signal window is represented as a solid-border rectangle.

Exclusive backgrounds coming from rare $B$ decays have been studied 
by means of large MC samples and 
only two channels have been found to give non-negligible contributions
within the signal window: \bpipi \ and \betapi.
Assuming the measured  \bpipi \ branching fraction,
$\mathrm{BF}(\bpipi) = 1.45 \pm 0.29 \times 10^{-6}$ ~\cite{hfag}, and 
the existing limit on the \betapi \ branching fraction,
$\mathrm{BF}(\betapi) < 2.5 \times 10^{-6}$ at 90\% CL~\cite{hfag}, 
0.09 \bpipi \ events and less than 0.06 \betapi \ events at 90\% CL are expected.
%
%

A two-dimensional extended unbinned maximum likelihood fit is performed on  
\deltae \ and \mbc \ to extract the signal yield.
The probability density functions (PDFs) for the signal are extracted 
from the MC simulation.
The photon energy resolution in the simulation is corrected to match
the resolution measured in a photon test beam~\cite{ECL}.
The signal PDFs are parametrized with a Crystal Ball
lineshape function~\cite{crystal} for \deltae \ and a double Gaussian  for \mbc.

For the continuum background, a linear
shape is assumed for \deltae, with the slope free to float in the fit, and
an ARGUS~\cite{argus} function for \mbc, with the slope parameter also free in the fit.
The exclusive backgrounds enter the fit with the normalization 
described above. They are parametrized
with a Gaussian PDF for \deltae \ and a double Gaussian
for \mbc.

\begin{figure}[t]
\includegraphics[width=0.48 \textwidth]{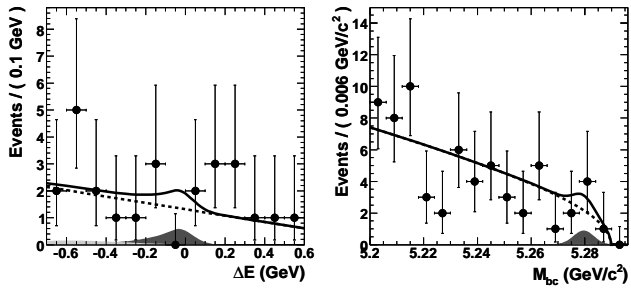}
\caption{Projections of the fit result on \deltae \
(with \mbc \ in its signal window) 
and on \mbc \ (with \deltae \ in its signal window).
The fitting curve (solid line) is 
plotted with data 
(circles with error bars, drawn as asymmetric Poisson confidence intervals) 
and the uds background (dashed line).
The filled regions represent the signal (dark shading)  and the
\bpipi \ and \betapi \ backgrounds (light shading).
\label{fig3}}
\end{figure}

The fit has four free parameters: two slopes and the numbers of events of 
the continuum background and of the signal.
It is performed within the \deltae \ range between $-0.7$  $\GeV$ and $0.6$ $\GeV$ 
and with \mbc \ greater than $ 5.2 \, \GeV/c^2$. The fit window is shown in 
Fig.~\ref{fig2} as a dashed rectangle.
The projections of the fit result on \deltae \ 
(with \mbc \ in its signal window) and on \mbc \ (with \deltae \ 
in its signal window) are shown in Fig.~\ref{fig3} as solid curves; 
the continuum
background is shown as dashed curves, the signal as the dark shaded
regions, and the exclusive backgrounds as the light shaded regions.

The signal yield is measured to be 
$ N_{\mathrm{sig}} = 1.8^{+3.5}_{-2.7}$, 
corresponding to a limit on the BF of $6.1 \times 10^{-7}$ at $90\%$ 
CL, obtained by integration
of the likelihood curve up to 90\% of its total area, and including
only the statistical uncertainty.

Several possible sources of systematic uncertainty are considered.
Uncertainties are included in the likelihood function
as additional parameters and then integrated over their respective
ranges by assuming Gaussian probability distributions.
The largest contribution is due to
the modelling of the signal shape, which depends on 
angular and energy resolutions of the calorimeter.
Uncertainties on these quantities, 
evaluated by studying samples of Bhabha and \eegg \ events,
have been propagated to the parameters of the signal PDFs
and to the fit result.
Other contributions are  
the uncertainties on the photon reconstruction efficiency, 
on event selection (LR and $\cos\theta^*$ requirements,
$\pi^0$ and $\eta$ mesons rejection), 
on the number of \bbar \ events, on background shapes,
and on the normalization of the exclusive backgrounds.
The separate contributions are summarized in Table~\ref{table:sys} as 
uncertainties on the signal yield. 
\begin{table}[thb]
\caption{ Summary of the main systematic sources, expressed as 
uncertainties on the fit signal yield.}
\label{table:sys}
\begin{tabular}
{l | c }
\hline 
Source                     &  Syst. unc. on $N_{\mathrm{sig}}$ \\
\hline
Signal shape                        &   $0.37$  \\
Photon rec. efficiency              &   $0.09$  \\
LR and $\cos\theta^*$ req.          &   $0.06$  \\
$\pi^0$ and $\eta$ vetoes           &   $0.05$  \\
Number of \bbar \ events            &   $0.03$  \\
Background shape and norm.          &   $0.02$  \\
\hline 
\end{tabular}
\end{table}

Inclusion of systematic uncertainties results in the following upper limit on the
BF: 
\begin{eqnarray}\nonumber 
\mathrm{BF}(\bgg) < 6.2 \times 10^{-7} \; \mathrm{at} \; 90\% \; \mathrm{CL} \, .
\end{eqnarray}
%
%
%

In conclusion, a search for the 
decay \bgg \ has been performed in 
104 \invfb \ of data with the Belle detector. No evidence of a
signal has been observed and a new upper limit 
has been set, corresponding
to an improvement of the previous limit of
about a factor of three.
%

\vspace*{\baselineskip}
We thank the KEKB group for the excellent operation of the
accelerator, the KEK cryogenics group for the efficient
operation of the solenoid, and the KEK computer group and
the National Institute of Informatics for valuable computing
and Super-SINET network support. We acknowledge support from
the Ministry of Education, Culture, Sports, Science, and
Technology of Japan and the Japan Society for the Promotion
of Science; the Australian Research Council and the
Australian Department of Education, Science and Training;
the National Science Foundation of China and the Knowledge 
Innovation Program of Chinese Academy of Sciencies under 
contract No.~10575109 and IHEP-U-503; the Department of 
Science and Technology of
India; the BK21 program of the Ministry of Education of
Korea, and the CHEP SRC program and Basic Reserch program 
(grant No. R01-2005-000-10089-0) of the Korea Science and
Engineering Foundation; the Polish State Committee for
Scientific Research under contract No.~2P03B 01324; the
Ministry of Science and Technology of the Russian
Federation; the Ministry of Higher Education, Science and 
Technology of the Republic of Slovenia;  the Swiss National 
Science Foundation; the National Science Council and
the Ministry of Education of Taiwan; and the U.S.\
Department of Energy.

\end{document}